\documentclass[pdflatex,sn-mathphys-num]{sn-jnl}

\usepackage{graphicx}%
\usepackage{multirow}%
\usepackage{amsmath,amssymb,amsfonts}%
\usepackage{amsthm}%
\usepackage{mathrsfs}%
\usepackage[title]{appendix}%
\usepackage{xcolor}%
\usepackage{textcomp}%
\usepackage{manyfoot}%
\usepackage{booktabs}%
\usepackage{algorithm}%
\usepackage{algorithmicx}%
\usepackage{algpseudocode}%
\usepackage{listings}%
\usepackage{natbib}

\theoremstyle{thmstyleone}%
\theoremstyle{thmstyletwo}%

\theoremstyle{thmstylethree}%

\begin{document}

\title[Article Title]{High-Endurance, Low-loss $\text{Sb}_2\text{Se}_3$ Optical Switches on Silicon Nitride using Transparent Conductive Heaters}


\author*[1]{\fnm{Xingshi} \sur{Yu}}\email{Xingshi.Yu@soton.ac.uk}

\author[1]{\fnm{Ipsita} \sur{Chakraborty}}\email{Ipsita.Chakraborty@soton.ac.uk}

\author[1]{\fnm{Isaac} \sur{Johnson}}\email{ij1g20@soton.ac.uk}

\author[2]{\fnm{Savvas I.} \sur{Raptis}}\email{sraptis@csd.auth.gr}

\author[1]{\fnm{Qianbin} \sur{Luo}}\email{Qb.Luo@soton.ac.uk}

\author[1]{\fnm{Thalia Dominguez} \sur{Bucio}}\email{T.Dominguez\_Bucio@soton.ac.uk}

\author[1]{\fnm{Elliot} \sur{Sandell}}\email{E.Sandell@soton.ac.uk}

\author[2]{\fnm{Chris} \sur{Vagionas}}\email{chvagion@csd.auth.gr}

\author[3]{\fnm{Ioannis} \sur{Zeimpekis}}\email{izk@soton.ac.uk}

\author[2]{\fnm{Amalia} \sur{Miliou}}\email{amiliou@csd.auth.gr}

\author[2]{\fnm{Nikos} \sur{Pleros}}\email{npleros@csd.auth.gr}

\author*[1]{\fnm{Frederic} \sur{Gardes}}\email{fg5v@ecs.soton.ac.uk}

\affil[1]{\orgdiv{Optoelectronics Research Centre (ORC)}, \orgname{University of Southampton }, \orgaddress{\street{University Road}, \city{Southampton}, \postcode{SO17 1BJ}, \state{Hampshire}, \country{United Kingdom}}}

\affil[2]{\orgdiv{Wireless and Photonic Systems and Networks (WinPhoS) research group}, \orgname{The Aristotle University of Thessaloniki}, \orgaddress{\street{3is Septemvriou Street, Agios Dimitrios}, \city{Thessaloniki}, \postcode{54124}, \country{Greece}}}

\affil[3]{\orgdiv{Electronics and Computer Science (ECS)}, \orgname{University of Southampton }, \orgaddress{\street{University Road}, \city{Southampton}, \postcode{SO17 1BJ}, \state{Hampshire}, \country{United Kingdom}}}

\abstract{We report an electrically actuated, low-loss non-volatile optical switch based on the phase-change material (PCM) $\mathrm{Sb}_2\mathrm{Se}_3$ integrated on a silicon nitride ($\mathrm{Si}_3\mathrm{N}_4$) platform. The device is fabricated using an 8-inch wafer-scale process flow, demonstrating the feasibility of scalable manufacturing for photonic integrated circuits (PICs). By employing transparent indium tin oxide (ITO) micro-heaters, reversible switching between the amorphous and crystalline states is achieved with an extinction ratio of 25~dB and an endurance exceeding 140 million switching cycles, establishing a new benchmark for non-volatile integrated photonic memory and reconfigurable architectures. Furthermore, multi-level operation beyond 6 bits can be repeatably demonstrated by tailoring the electrical pulse widths, enabling precise control of the optical phase. These results highlight a scalable and energy-efficient platform for high-density programmable and non-volatile photonic integrated systems.}


\maketitle
\section*{Main}
Silicon nitride ($\text{Si}_3\text{N}_4$) provides a low-loss platform for integrating photonic components. Its low thermo-optic coefficient enhances thermal stability, while its moderate refractive index contrast improves fabrication tolerance \cite{gardes2022review, xiang2022silicon}. These features support scalable, high-yield photonic integrated circuits (PICs) for next-generation data centers and high-performance computing \cite{wan2026integrating,shekhar2024roadmapping,soref2006past}. Central to these systems of applications are phase shifters, enabling dynamic signal routing, reconfigurable multiplexing, and programmable photonic functionalities \cite{xu2025progress, serafino2020high, miller2009device}. Conventional implementations based on thermo-optic or carrier-induced plasma dispersion effects are inherently volatile, requiring continuous electrical bias to maintain a programmed state \cite{parra2024silicon, zhang2025all, zhang2023harnessing, zabelich2024}. 

The integration of non-volatile phase-change materials (PCMs) provides a promising route to overcome this limitation \cite{prabhathan2023roadmap}. $\text{Sb}_2\text{Se}_3$ exhibits low absorption while providing reversible refractive index changes between amorphous and crystalline phases, enabling zero-static-power optical switching \cite{fang2022ultra, taghinejad2021ito, delaney2020new, faneca2021towards}. These transitions can be triggered optically or electrically, supporting applications in photonic memory, neuromorphic computing, and optical field-programmable gate arrays (FPGA) \cite{wang2021electrical, guo2026variable, zarei2026ultracompact, xu2025back}. On-chip resistive heaters allow compact integration and compatibility with established electronic packaging processes. Compared with conventional metallic heaters, transparent conductive materials with lower refractive index, such as indium tin oxide (ITO), are widely adopted to further minimize optical loss arising from substantial parasitic metallic absorption\cite{xia2024ultrahigh}.

\begin{figure}[htbp]
\centering\includegraphics[width=13cm]{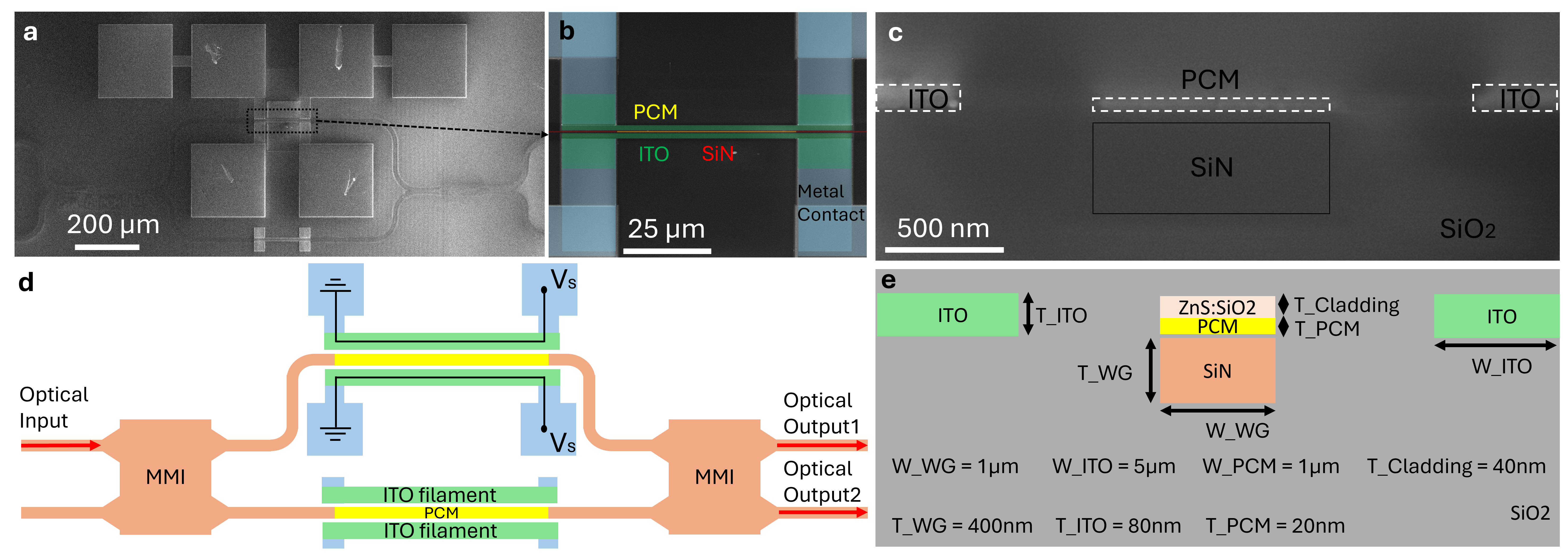}
\caption{Device design and structural characterization of the MZI with an integrated PCM phase shifter.
(a) Top-view SEM image of the fabricated MZI device.
(b) Enlarged SEM image of the phase shifter integrated on one arm of the MZI.
(c) Cross-sectional SEM image of the PCM phase-shifter active region.
(d) Schematic top view of the overall device layout.
(e) Schematic cross-section of the phase-shifter active region.}
\label{design}
\end{figure}

We present a high-endurance non-volatile optical switch based on the heterogeneous integration of $\text{Sb}_2\text{Se}_3$ on a $\text{Si}_3\text{N}_4$ platform. We utilize indium tin oxide (ITO) as a transparent resistive micro-heater to trigger the phase transition. This configuration enables low-loss, efficient, localized Joule heating, facilitating reversible switching between the amorphous and crystalline states of the $\text{Sb}_2\text{Se}_3$ film \cite{taghinejad2021ito, xia2024ultrahigh}. The non-volatile photonic switch is designed using a Mach–Zehnder interferometer (MZI) configuration. In Fig.~\ref{design}, the photonic circuitry utilizes $\text{Si}_3\text{N}_4$ waveguides with width and thickness of 1~$\mu$m $\times$ 400~nm respectively, optimized for single-mode operation of the fundamental transverse electric (TE) mode at telecommunication wavelengths \cite{bucio2019silicon}. The active switching mechanism is based on the phase transition of a 20~nm thick $\text{Sb}_2\text{Se}_3$ film \cite{miyatake2025low, gutierrez2022interlaboratory}. This PCM is deposited on a 20~nm thick $\text{SiO}_2$ optical spacer, which serves as a critical buffer layer to minimize surface scattering and protect the waveguide core from interfacial diffusion during thermal processes \cite{chen2024silatrane, trempa2020impact}. To suppress oxidation of PCM and improve thermal stability during cycling, a 40~nm $\text{ZnS:SiO}_2$ (20\%:80\%) cladding layer is used \cite{laprais2024reversible}. The relatively low thermal conductivity helps balance the thermal conditions required for amorphization and crystallization \cite{teo2023capping}. A key design feature is the 300~nm lateral gap between the $\text{Si}_3\text{N}_4$ waveguide core and the integrated ITO microheater filaments. Electro-thermal and optical co-simulations indicate that, at this separation, the evanescent field of the fundamental TE mode decays sufficiently, limiting optical absorption from the ITO elements to below $0.005\,\mathrm{dB}/\mu\mathrm{m}$, while maintaining efficient heat transfer. Based on four-probe measurements, the sheet resistance ($R_{\text{sh}}$) of the in-house deposited ITO layer was ascertained to be $17~\Omega/\text{sq}$. To ensure compatibility with standard co-packaging optics (CPO) drivers, the operating voltage was maintained below $\sim$10~V by fixing the filament width at $5~\mu$m~\cite{chowdhury2025electronic}, while scaling its length with the PCM region. The detailed fabrication processes are shown in Supplementary Section 1.

\section*{Results}\label{sec2}
\subsection*{Electrical Characterization of ITO Microheaters}\label{subsec1}
\begin{figure}[htbp]
\centering\includegraphics[width=13cm]{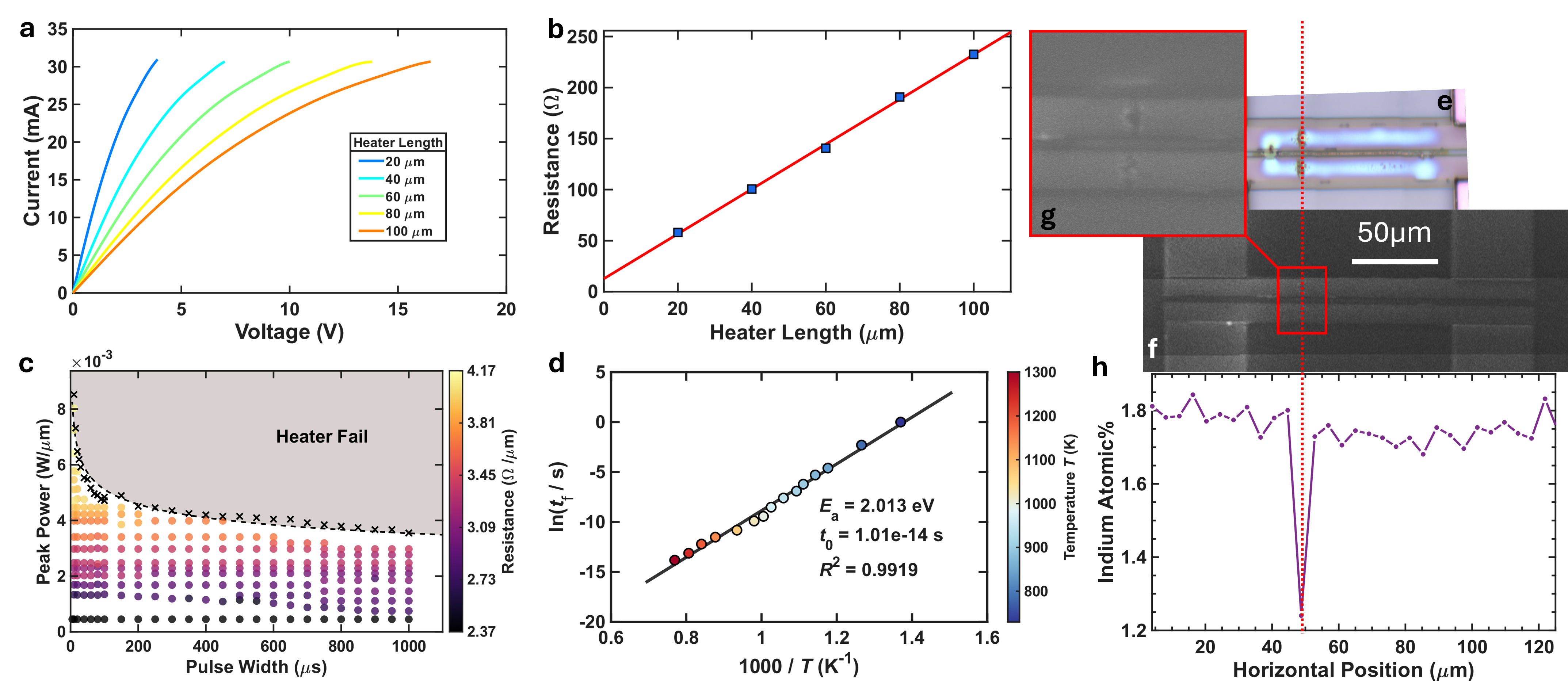}
\caption{Current-voltage characteristics of ITO filaments with varying lengths
(a) Current-voltage characteristics of ITO filaments with varying lengths.
(b) Ohmic resistance of ITO filaments with varying lengths (measured at 0.1~V per 20~$\mu$m)
(c) Normalized resistance of a pair of ITO filaments versus pulse cycles for different pulse widths and corresponding pulse peak powers.
(d) Arrhenius fitting of the time–temperature superposition for ITO failure..
(e) Optical microscopy image of the fractured ITO filament.
(f) Top-view SEM image of the PCM phase shifter containing the broken ITO filament.
(g) Magnified SEM image of the fracture region of the ITO filament.
(h) EDS line-scan analysis across the ITO filament. The x-axis represents the scan position along the filament from left to right (as indicated above), while the y-axis corresponds to the detected indium atomic density.
}
\label{ITO_IV}
\end{figure}

For all experiments and device measurements, two ITO filaments were characterized in parallel. The voltage–current characteristics of micro-heaters with varying length were recorded, as shown in Fig.~\ref{ITO_IV} (a), to evaluate their resistance stability and maximum power-handling capability. Notably, linearity was observed near zero bias, indicating the absence of Schottky barriers and confirming the integrity of the ITO–metal interfaces \cite{hwang2005novel}. For all heater lengths, device failure occurred at a visually detectable rate once the total current reached ~30 mA under DC bias. The $V$–$I$ characteristics of the ITO filaments exhibit a gradual increase in resistance with applied voltage, reflecting Joule heating. To extract the intrinsic resistance at room temperature, we employed a transmission line method measurement using a low bias of 0.1~V per 20~$\mu$m, yielding 2.37~$\Omega/\mu$m and 23.8~$\Omega$ per contact pad (Fig.~\ref{ITO_IV} (b)) \cite{berger1972models}. The resistivity is lower than the four-probe estimate due to conductivity enhancement from annealing \cite{Lotkov2022}. To investigate the dependence of filament resistivity on electrical stress, we applied pulses of varying width and amplitude to filaments of different lengths, recording the resulting current response. The measurement setup is shown in Supplementary Section 2. From these measurements, a map of the normalized resistance as a function of pulse width and peak electrical power was constructed (Fig.~\ref{ITO_IV} (c)), providing insight into the Joule-heating-induced resistance changes.
The resistivity exhibits a clear monotonic increase with peak electrical power in the two-filament heater system, consistent with a positive temperature coefficient of resistance. At elevated temperatures, enhanced electron–phonon scattering reduces carrier mobility, leading to increased resistivity \cite{kikuchi2000phonon}. At a fixed power density, the failure threshold strongly depends on the pulse width, as longer pulses impose a larger cumulative thermal load on the heaters within a given time window. 

We further analyzed the voltage-resistance data located at the conductive filament failure boundary. Measurements were performed under direct current conditions as well as under pulsed operation with a minimum pulse width of 1~$\mu$s. In addition, multiphysics simulations were carried out using \textit{COMSOL Multiphysics} to extract the corresponding failure temperature of ITO (detail about the model is in Supplementary section 3). The obtained data were subsequently fitted using the Arrhenius analysis model

\begin{equation}
\ln t = \ln t_0 + \frac{E_a}{k_B} \cdot \frac{1}{T}
\end{equation}
 
The fitting results are presented in Fig.~\ref{ITO_IV} (d). $k_B$ denotes the Boltzmann constant, which links energy and temperature. The Arrhenius fit yields an activation energy of $E_a = 2.01$ eV and a pre-exponential factor of $t_0 = 1.01 \times 10^{-14}$ s, with an excellent coefficient of determination $R^2 = 0.9919$. In statistical terms, $R^2$ quantifies the proportion of the variance in the observed data that is captured by the fitted model. The activation energy of approximately $2.01$ eV is consistent with the energy scale of thermally activated point defect formation \cite{hou2018defect, liu2014thermodynamics}, suggesting the failure is governed by thermally driven lattice rearrangement. Furthermore, the extracted pre-exponential factor is on the order of $10^{-14}$ s, which is comparable to the characteristic timescale of atomic vibrational processes ($\sim 10^{-14}$--$10^{-13}$ s) \cite{nitzan2024chemical}, confirming the physical consistency of the fitted kinetic parameters.  

Fig.~\ref{ITO_IV} (e) shows optical micro-graphs of failed heaters, where pronounced breakdown along each filament is clearly observed. Energy-dispersive X-ray spectroscopy (EDS) performed at the fractured regions reveals a significantly reduced indium atomic density compared with intact sections of the heaters (in Fig.~\ref{ITO_IV} (h)). Under high-bias conditions, the micro-heaters reach their operational limit as excessive temperature. The resulting atomic displacement promotes the formation of localized conductive paths, causing severe current crowding and runaway thermal loading, ultimately leading to catastrophic breakdown of the ITO layer \cite{leung2013study, chao1996failure}. 

\subsection*{Operational Boundaries}\label{subsec2}
\begin{figure}[htbp]
\centering\includegraphics[width=13cm]{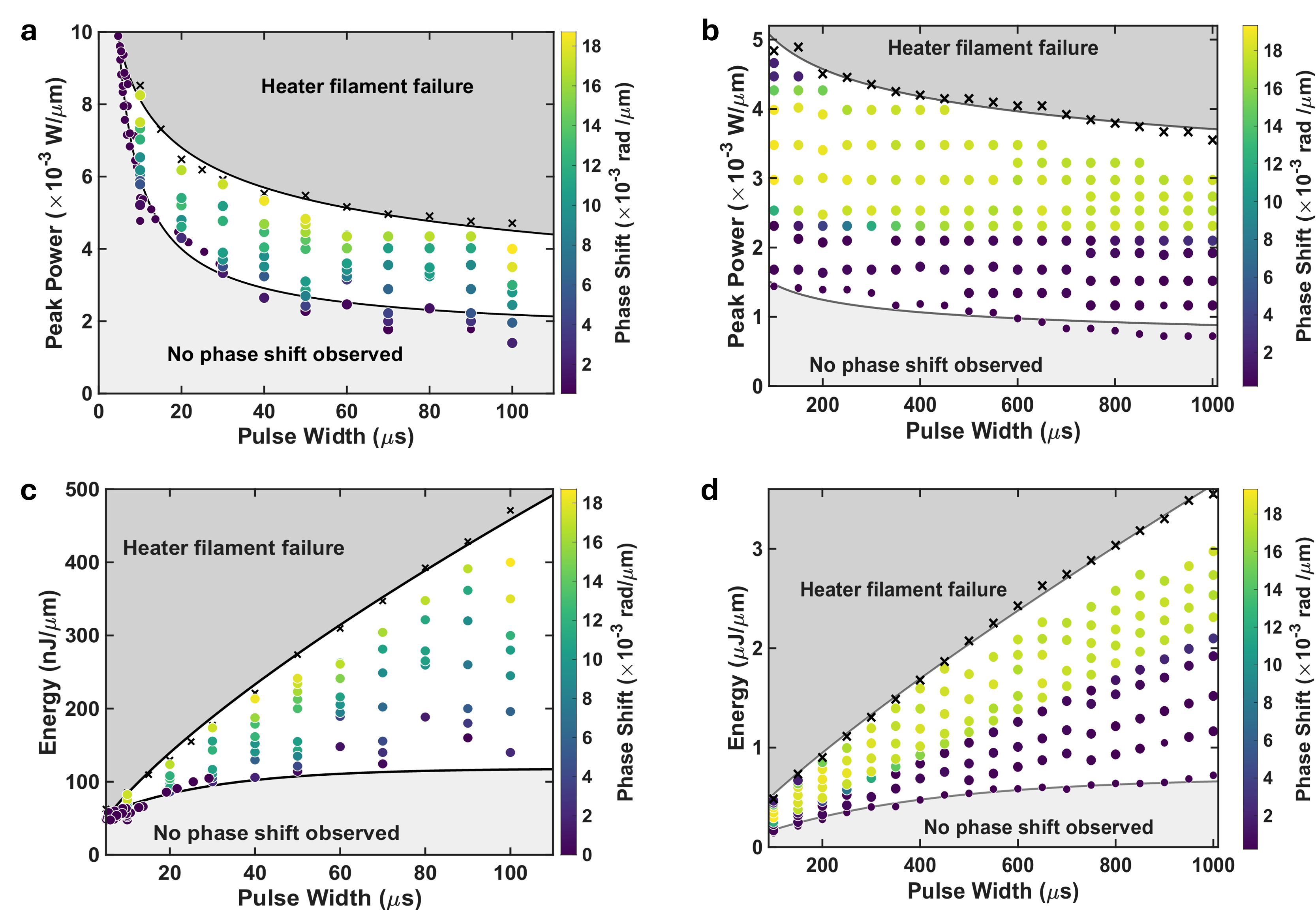}
\caption{Operational switching window of a normalized length PCM switch.
(a) Operational switching window of the ITO heaters for set.
(b) Operational switching window of the ITO heaters for reset.
(c) Switching energy trend for set.
(d) Switching energy trend for reset.}
\label{Mapping}
\end{figure}

Based on the preceding characterization of the ITO heaters, the filaments are expected to fail beyond a critical electrical stress. Accordingly, we investigate the phase shift as a function of peak electrical power during the initial cycling, using an 80~$\mu\mathrm{m}$-long PCM phase shifter integrated in a MZI. Supplementary Sections 4–6 show that both C-band and O-band devices exhibit a consistent linear dependence of phase shift on PCM length. The operational envelope of the device was mapped by systematically varying the set-pulse (amorphization) power and pulse width (under 100~$\mu$s) to identify the limits of non-volatile phase modulation. For each data point, the same reset (crystallization) peak power and width of the pulse (0.0035~W/$\mu$m, 600~$\mu$s) were applied to ensure consistent conditions. At low peak power of set-pulse, no detectable phase transition was observed. The minimum power density required to induce a phase change increases as the pulse width decreases, reflecting insufficient thermal accumulation under shorter pulses. This lower boundary of the phase-transition regime is indicated by the upper boundary of the no-shift region in Fig.~\ref{Mapping} (a). Conversely, at excessively high peak power of set-pulses, the ITO heater undergoes rapid electrical failure due to resistive breakdown, defining the lower boundary of the failure region. Together, these two boundaries delineate the effective set window of the device. Notably, although both boundaries exhibit similar trends with pulse width, the operational window narrows dramatically as the pulse width approaches 5~$\mu$s and eventually collapses. In this regime, irreversible ITO heater failure occurs before sufficient thermal energy can be delivered to induce phase shift, thereby precluding any viable set operation. Within the accessible working window, larger pulse power at a fixed pulse width results in increased phase shift, reaching a maximum modulation efficiency of $0.018~\mathrm{rad}/\mu$m. 

Furthermore, the operational boundaries of the reset process were investigated, as shown in Fig.~\ref{Mapping}(b). Increasing the pulse duration beyond 100~$\mu$s requires lower electrical power to reach the reset temperature of the PCM, due to enhanced thermal dissipation over longer timescales. For each data point, a fixed set pulse (20~$\mu$s, 0.006~W/$\mu$m) was applied to return the device to a reference state. A stable working window is observed for the reset operation. Compared with the set process, the boundary evolves more gradually with pulse width and does not exhibit closure within the explored parameter range. For pulse widths exceeding 250~$\mu$s, all pulses below the failure threshold consistently achieve full reset. In contrast, for pulse widths shorter than 200~$\mu$s, increasing the power density does not significantly improve the reset response, which is attributed to partial crystallization induced by short, high-power pulses. These observations indicate a trade-off between pulse width, energy consumption, and reliable switching. Energy analysis (Fig.~\ref{Mapping}(c,d)) shows that shorter pulses reduce the total energy required for a given phase shift, but impose stricter constraints on reliable operation. 

Based on these results, long-term reliability experiments were performed using a set pulse of 20~$\mu$s at 0.006~W/$\mu$m and a reset pulse of 600~$\mu$s at 0.0035~W/$\mu$m, ensuring operation within the stable working window while maintaining low energy consumption and repeatable phase modulation.

\subsection*{Endurance}\label{subsec3}
\begin{figure}[htbp]
\centering\includegraphics[width=13cm]{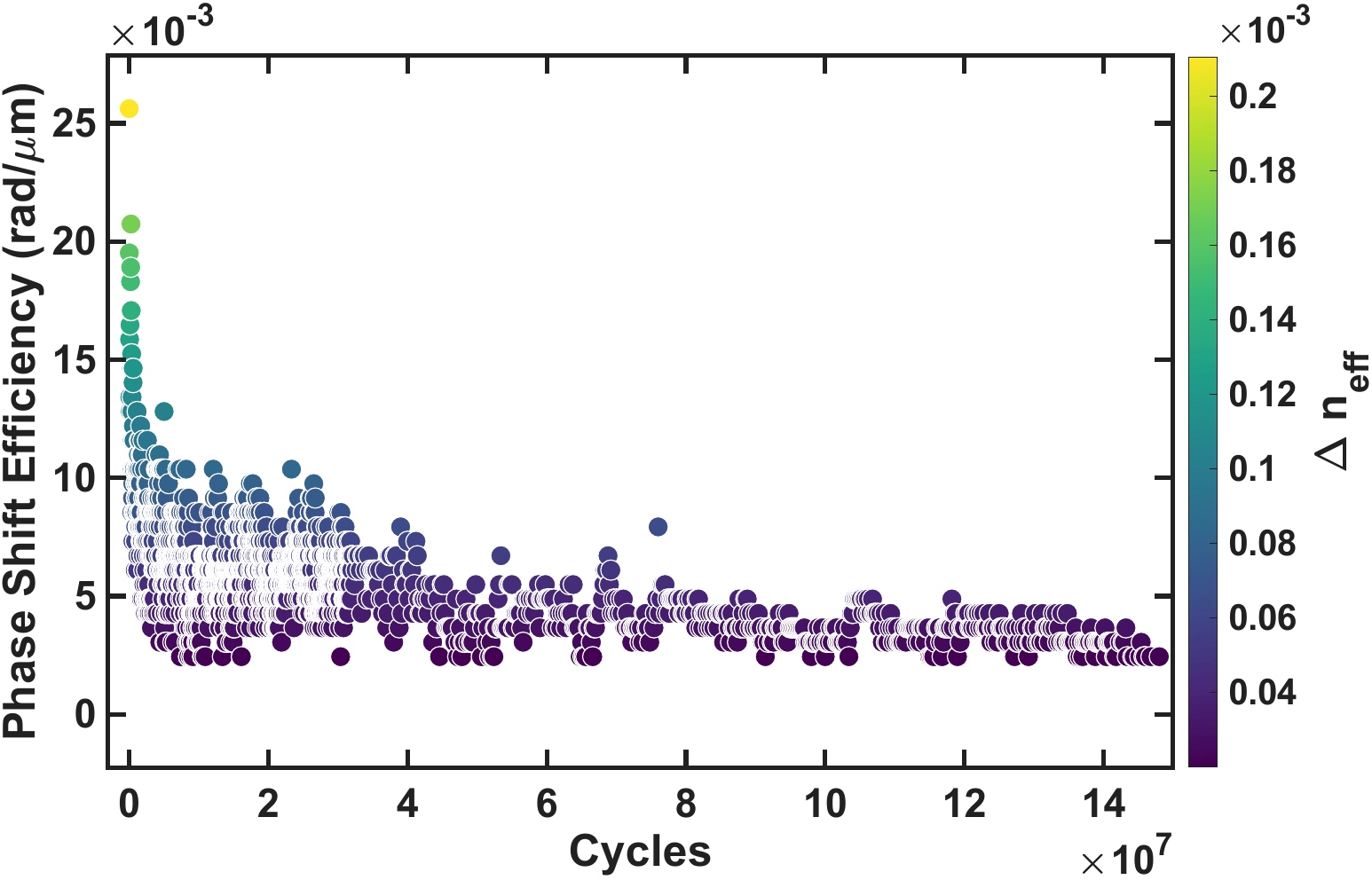}
\caption{Endurance cycling test results showing the extinction ratio over 140 million cycles.}
\label{100 million}
\end{figure}

In this section, a C-band MZI with 30~$\mu\mathrm{m}$ PCM phase shifters is used to assess long-term reliability. During the non-volatile switching cycles, the set pulse is $0.006~\mathrm{W}/\mu\mathrm{m}$ with a duration of 20~$\mu$s, and the reset pulse is $0.0035~\mathrm{W}/\mu\mathrm{m}$ with a duration of 600~$\mu$s. Due to the limitation of maximum output voltage stated in Supplementary section 4. An MZI with 30-$\mu$m-long $\mathrm{Sb}_2\mathrm{Se}_3$ phase shifters was subjected to continuing set and reset cycles, during which the corresponding transmission shift were monitored in real time. Fig.~\ref{100 million} illustrates the normalized phase shift efficiency, and the extracted effective refractive index change during repeated cycling. The phase shift efficiency stabilizes at approximately $0.004\,\mathrm{rad/\mu m}$ after approximately 1 million cycles. This initial stabilization process is attributed to interfacial stress and material fatigue within the $\text{SiO}_2$ cladding \cite{constantin2022new, schrauwen2008trimming, xi2013modeling}. The resistance of the ITO heater exhibits a significant increase after approximately 40 million switching cycles. To maintain the required electrical peak power ($0.006~\mathrm{W}/\mu\mathrm{m}$ for set, $0.0035~\mathrm{W}/\mu\mathrm{m}$ for reset) delivered to the heater, the voltage of both the set and reset pulses is increased accordingly (see Supplementary section 7), ensuring a consistent level of phase shift. The $\mathrm{Sb}_2\mathrm{Se}_3$-based phase shifters exhibit robust and repeatable operation for up to $140$~million switching cycles. Beyond $140$~million cycles, the phase change progressively decreased toward zero at a rate of $0.001\,\mathrm{rad/\mu m}$ per 5 million cycles, indicating device failure associated with electrical breakdown of the ITO heaters. 

In Table~\ref{tab:PCM_phase_shifters}, we summarize representative studies on non-volatile photonic phase shifters. Our results demonstrate the endurance and bit precision that significantly exceed those of previous reports based on PCM technologies. Moreover, this work substantially narrows the performance gap with state-of-the-art volatile technologies used for digital-analog hybrid matrix multiplication in image processing and deep learning applications \cite{bai2023microcomb, zhang2022silicon}. This improvement can be attributed to the careful control of the heater operating boundaries, which ensures stable and reliable device operation over extended cycling. Compared with state-of-the-art low-loss $\mathrm{Sb}_2\mathrm{Se}_3$ photonic phase shifters, which report a large intrinsic refractive index contrast between the fully crystalline and fully set phases at telecommunication wavelengths \cite{delaney2020new,delaney2021nonvolatile,rios2022ultra}, the effective refractive index change and phase shift efficiency we achieved is noticeably smaller. This indicates that the electrically programmed set state corresponds to an incompletely amorphized configuration rather than a fully amorphous phase.

\begin{table*}[t]
\centering
\caption{Comparison of non-volatile photonic memory demonstrations.}

\resizebox{\textwidth}{!}{
\begin{tabular}{lccccccc}
\hline
\toprule
Ref. &
Technology / Material &
1 bit endurance &
Multi-bit number and cycles &
1 bit Switching Time &
1 bit Switching Energy &
Shifting Efficiency (rad\,$\mu$m$^{-1}$) &
ER (dB) \\
\midrule
\hline
\cite{hu2025nonvolatile} & MEMS & $10^{6}$ & N/A & 2.2 µs / 7.4 µs & 1 pJ & N/A & 44.4 \\

\cite{geler2022ferroelectric} & Ferroelectric & 300 & 3bit, 300 & 1 µs & 4.6 pJ & 0.003 & 12 \\

\cite{tossoun2024high} & Memristor (GaAs-Al$_2$O$_3$-Si) & 1,000 & 1.5-bit, 1,000 & 0.3 ns / 0.9 ns & 150 fJ / 360 fJ & 0.0282 & 14 \\

\cite{li2025pzt} & Memristor PZT & $>$100000 & 2.32-bit, 20 & 2.5 ns & 12.3 pJ & 0.003 & 19 \\

\cite{pintus2025integrated} & Magneto-optic Ce:YIG & $2.4\times10^9$ & 3.5-bit, 1 & 1 ns & 143 fJ & 0.0024 & 16.2 \\

\cite{fang2022ultra} & Sb$_2$Se$_3$, Graphene heater & 1500 & 3.5bit, 1 & 408 ns / 220 µs & 5.55 nJ / 753 nJ & 0.00035 & 3 \\

\cite{zhou2023memory} & GST, Doped Si & 100 & 3.5bit, 3 & 50 ns / 200 ns & 8.8 nJ / 6.9 nJ & N/A & 4.13 \\

\cite{yang2023non} & Sb$_2$Se$_3$, Doped Si & 10,000 & 6bit, 1 & 135 ns / 20 µs & 105 nJ / 1.2 µJ & 0.215 & 20 \\

\cite{meng2023electrical} & GSSe, W/Ti & 500,000 & 4 bit,1 & 2 µs / 20 µs & 400 nJ & N/A & 6 \\

\cite{xia2024ultrahigh} & Sn-GST, ITO & 30,000 & 3.87 bit,16 & 5 µs / 50 µs & N/A & N/A & 50 \\

\cite{zhang2023nonvolatile} & GST, ITO & 3 & 6 bit,1 & 1.1 µs / 60 µs & 81 nJ / 1.96 µJ & 0.04 & 30 \\

\cite{shoaa2025stable} & Sb$_2$Se$_3$, Doped Si & 7900 & NA & 1.5 µs / 300 µs & 0.54 µJ / 19.2 µJ & 0.067 & 12 \\

This work & Sb$_2$Se$_3$, ITO & $1.4\times10^8$ & 6 bit, 10 & 20 µs/ 600 µs & 3.6 µJ / 630 µJ & 0.004 & 29 \\
\hline
\bottomrule
\end{tabular}
}
\label{tab:PCM_phase_shifters}
\end{table*}

\subsection*{Multi-level Modulation}\label{subsec4}
\begin{figure}[htbp]
\centering\includegraphics[width=13cm]{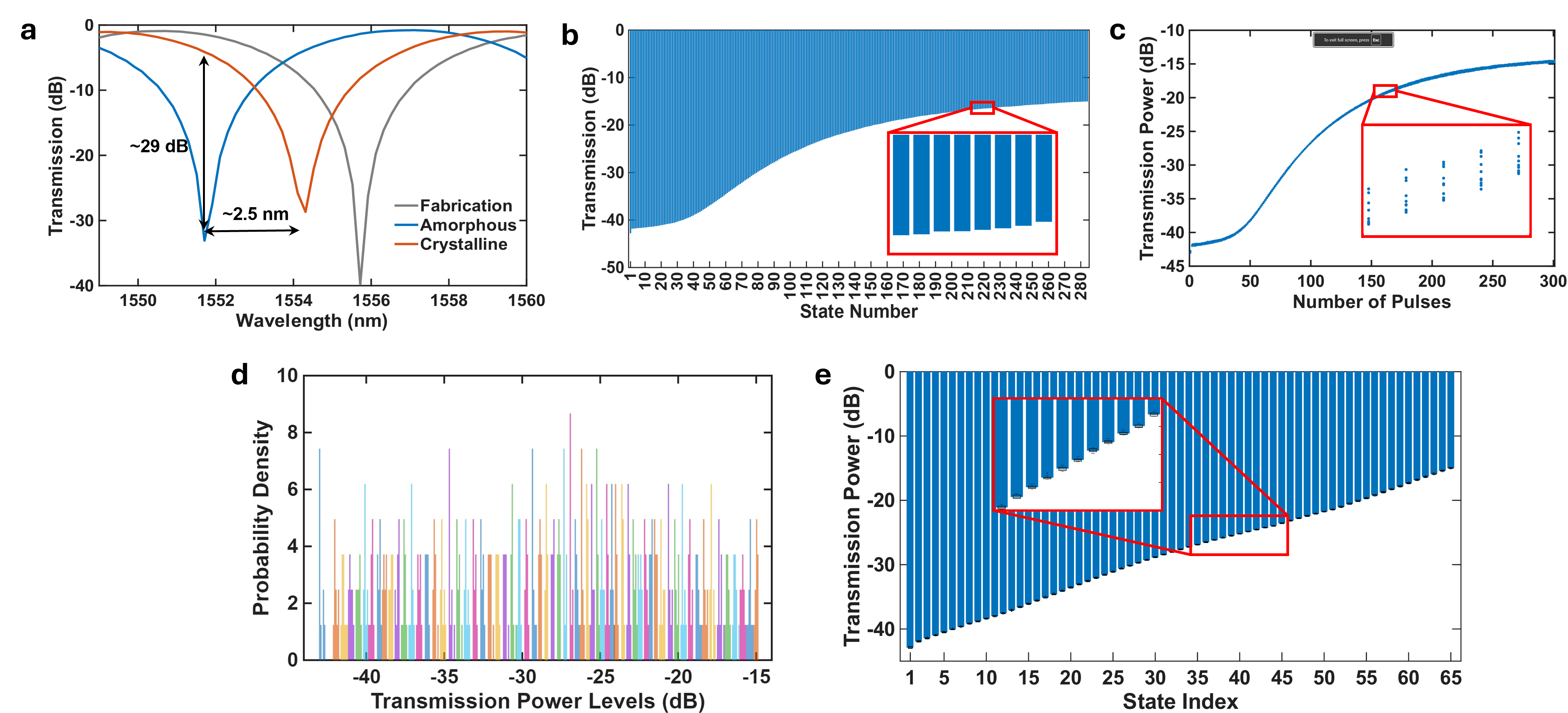}
\caption{Multi-level pulse-width modulation based on a 80~$\mu$m length PCM switch.
(a) Initial measurement for Amorphous and crystalline states’ spectrum.
(b) Gradual crystallization of the PCM using low duration crystallization pulses.
(c) Multiple repetition of gradual crystallization, capturing the transmission power level.
(d) Distribution of power across 65 states, including the representation of the 65 discrete levels with overlaid boxplot analysis.
(e) The probability density functions for all discrete states, with standard deviation values varying from 0.0566 to 0.1265.
}
\label{Multi}
\end{figure}

To explore the achievable bit precision in PCM-based MZIs, a device incorporating an 80~$\mu$m-long PCM section in one arm was characterized. A set of $15.4~\mathrm{V}$, $40~\mu\mathrm{s}$ pulses was used for amorphization, while $10.4~\mathrm{V}$, $600~\mu\mathrm{s}$ pulses were applied for crystallization. The measured transmission spectrum over a single free spectral range (1550--1560~nm) is shown in Fig.~\ref{Multi}(a), comparing the initial, amorphous, and crystalline states. An extinction ratio of up to 29~dB is achieved at 1551.75~nm, with a maximum wavelength shift of 2.5~nm between states.

Multi-level operation was further demonstrated via gradual crystallization. Starting from the amorphous state, a sequence of 300 identical pulses ($10.4~\mathrm{V}$, $80~\mu\mathrm{s}$) was applied while monitoring the transmission at 1551.75~nm. The transmission increases monotonically with pulse number, indicating progressive phase transitions in the PCM. As shown in Fig.~\ref{Multi}(b), this process yields 284 distinguishable transmission levels spanning an aggregate dynamic range of 27~dB, corresponding to an average step size of 0.095~dB between adjacent levels.

Repeatability was evaluated over ten independent switching cycles. In each cycle, the PCM was reinitialized to the amorphous state using $15.4~\mathrm{V}$, $40~\mu\mathrm{s}$ pulses, followed by the same crystallization sequence. The resulting transmission traces, presented in Fig.~\ref{Multi}(c), exhibit strong overlap, confirming stable and repeatable operation across cycles.

To quantify the statistical distribution of the transmission states, each level is modeled as a Gaussian distribution with mean $\mu_n$ and standard deviation $\sigma_n$. The probability density function of the $n$-th state is given by
\begin{equation}
\text{PDF}_n =
\frac{1}{\sigma_n \sqrt{2\pi}}
\exp\left[-\frac{(P_n - \mu_n)^2}{2\sigma_n^2}\right].
\end{equation}

Based on the overlap of adjacent distributions, the number of reliably distinguishable states is determined to be 65, corresponding to a symbol error rate (SER) of $1.2\times10^{-2}$. The distributions of all states are summarized in Fig.~\ref{Multi}(d,e), with standard deviations ranging from 0.0566 to 0.1265~dB, confirming the robustness of multi-level operation.

\section*{Discussion}
We demonstrate a scalable, electrically driven, multi-level photonic computing platform based on a non-volatile, electrically reconfigurable phase shifter employing the low-loss phase-change material $\mathrm{Sb}_2\mathrm{Se}_3$ integrated with ITO micro-heaters on an 8-inch $\mathrm{Si}_3\mathrm{N}_4$ fabrication platform. 

The device sustains approximately $1.4\times10^8$ reversible 1-bit switching operations, achieving a phase-shifting efficiency of $0.004\,\mathrm{rad/\mu m}$ and an effective refractive index contrast of $\Delta n_\mathrm{eff} \approx 0.003$, as measured experimentally. This operating point ($0.004\,\mathrm{rad/\mu m}$) corresponds to conditions well within the stable switching window of Fig.~\ref{Mapping}(a), prioritising endurance over maximum phase-shift amplitude.

The reported endurance is derived from full set/reset cyclings at elevated temperature using an Arrhenius model, representing a worst-case operating condition. In practical applications, devices predominantly undergo non-destructive read operations at low optical power, interspersed with relatively infrequent write events. Consequently, the effective lifetime under realistic partial-cycling conditions is expected to exceed the full-cycle estimate. For application scenarios such as optical circuit switching (OCS) in data centers, programmable photonic FPGAs, and synaptic weight updates in optical neural networks (ONNs), the typical switching demand is on the order of $10^{4}$–$10^{5}$ write operations per day. This corresponds to a projected device lifetime exceeding $\sim 4$ years, and substantially longer under read-dominated workloads, highlighting the suitability of the device for long-term system-level deployment.

The measurements indicate that device endurance is presently limited by heater failure rather than intrinsic PCM degradation. Replacing the ITO heater with more thermally robust transparent conductors (e.g., IZO, AZO, or Nb-doped TiO$_2$) may therefore decouple the endurance ceiling from the PCM and potentially extend the switching lifetime beyond $10^9$ cycles. In addition, such materials could sustain higher peak power, enabling faster thermal transients and enhanced quench rates, which may improve amorphization efficiency and yield larger phase-shift contrast \cite{loke2012breaking,zalden2019femtosecond}.

Multi-level operation enabled by pulse-width modulation allows continuous and deterministic tuning of optical transmission. In our device, 284 optically resolvable levels collapse to 65 repeatable levels, corresponding to a measured state error rate (SER) of $1.2 \times 10^{-2}$. This limitation is consistent with stochastic nucleation during partial crystallization of the PCM, which fundamentally constrains the number of reliably distinguishable states. Within the Johnson–Mehl–Avrami–Kolmogorov (JMAK) framework, the intermediate crystalline fraction for a given pulse energy is governed by Poisson-distributed nucleation events, leading to shot-to-shot variations in the effective refractive index. 

Further reduction of SER can be achieved by increasing the phase-change efficiency to produce larger effective index shifts per level, thereby improving level separation, as well as through optimization of the electrical driving scheme. Currently, multi-level writing is performed in an open-loop fashion, applying a fixed pulse sequence without real-time feedback. While this allows high-speed programming, it is inherently limited by stochastic nucleation. Incorporating a closed-loop feedback scheme, where the optical output is monitored during writing and used to adjust subsequent pulses (optical servo), would improve multilevel precision at the cost of increased write latency. This trade-off between speed and fidelity positions our work within the context of emerging PCM photonic memory architectures and provides a clear pathway for future optimization of device performance.

Alternatively, the phase shift can be increased by scaling the PCM length. The present MZI incorporates $80~\mu\mathrm{m}$ PCM phase shifters, corresponding to a 29~dB ER, while the structure supports ER values exceeding 33~dB (Supplementary Section 5). However, increasing the PCM length also enlarges thermal non-uniformity and the stochastic nucleation volume, leading to enhanced shot-to-shot variability. This trade-off implies the existence of an optimal PCM length that maximizes the number of reliable levels for a given SER. If this optimal length is shorter than that required for a full phase-amplitude interference shift, a cascaded (multi-stage) PCM architecture provides a viable route to simultaneously achieve high optical resolution and low SER.

We evaluate the energy efficiency of our multi-level PCM photonic switch by computing the energy per programmed optical state. For our current 6-bit operation (65 repeatable levels), the set (amorphization) pulse requires approximately $19.2~\mu\mathrm{J}$, while each intermediate reset (crystallization) pulse consumes $22.4~\mu\mathrm{J}$ per level. This corresponds to an energy-per-level figure of merit, allowing direct comparison to conventional electronic memories: Dynamic random-access memory (DRAM) operates at $\sim 10~\mathrm{fJ/bit}$, and Flash at $\sim 10~\mathrm{pJ/bit}$, highlighting the orders-of-magnitude higher energy cost typical of PCM photonic switches. To further improve energy efficiency, multi-level reset pulses can employ the same amplitude as the set pulses while reducing the pulse width to the $\sim 10~\mu\mathrm{s}$ scale. This approach delivers the same peak temperature necessary for crystallization, enabling the desired phase-change effect with approximately 100-times lower energy consumption per pulse. This approach situates the device within the broader memory landscape and provides a framework for systematic energy-performance comparisons between PCM photonic devices and conventional electronic memories.

The results in this work provide a comprehensive framework for the development of energy-efficient, broadband non-volatile photonic systems, paving the way for next-generation optical interconnects and hardware-efficient neural networks.

\section*{Data availability}
The data that support the findings of this study are available from the corresponding authors.

\section*{Code availability}
The code that support the findings of this study are available from the corresponding authors.

\bibliography{sn-bibliography}

\section*{Acknowledgements}

Moved to cover letter for Double-anonymised

\section*{Author contributions}

Moved to cover letter for Double-anonymised 

\section*{Competing interests}
The authors declare no competing interests.

\section*{Supplementary}
Supplementary Information (download PDF).

\end{document}


\title[Article Title]{Supplementary Information: High-Endurance, Low-loss $\text{Sb}_2\text{Se}_3$ Optical Switches on Silicon Nitride using Transparent Conductive Heaters}


\author*[1]{\fnm{Xingshi} \sur{Yu}}\email{Xingshi.Yu@soton.ac.uk}

\author[1]{\fnm{Ipsita} \sur{Chakraborty}}\email{Ipsita.Chakraborty@soton.ac.uk}

\author[1]{\fnm{Isaac} \sur{Johnson}}\email{ij1g20@soton.ac.uk}

\author[2]{\fnm{Savvas I.} \sur{Raptis}}\email{sraptis@csd.auth.gr}

\author[1]{\fnm{Qianbin} \sur{Luo}}\email{Qb.Luo@soton.ac.uk}

\author[1]{\fnm{Thalia Dominguez} \sur{Bucio}}\email{T.Dominguez\_Bucio@soton.ac.uk}

\author[1]{\fnm{Elliot} \sur{Sandell}}\email{E.Sandell@soton.ac.uk}

\author[2]{\fnm{Chris} \sur{Vagionas}}\email{chvagion@csd.auth.gr}

\author[3]{\fnm{Ioannis} \sur{Zeimpekis}}\email{izk@soton.ac.uk}

\author[2]{\fnm{Amalia} \sur{Miliou}}\email{amiliou@csd.auth.gr}

\author[2]{\fnm{Nikos} \sur{Pleros}}\email{npleros@csd.auth.gr}

\author*[1]{\fnm{Frederic} \sur{Gardes}}\email{fg5v@ecs.soton.ac.uk}

\affil[1]{\orgdiv{Optoelectronics Research Centre (ORC)}, \orgname{University of Southampton }, \orgaddress{\street{University Road}, \city{Southampton}, \postcode{SO17 1BJ}, \state{Hampshire}, \country{United Kingdom}}}

\affil[2]{\orgdiv{Wireless and Photonic Systems and Networks (WinPhoS) research group}, \orgname{The Aristotle University of Thessaloniki}, \orgaddress{\street{3is Septemvriou Street, Agios Dimitrios}, \city{Thessaloniki}, \postcode{54124}, \country{Greece}}}

\affil[3]{\orgdiv{Electronics and Computer Science (ECS)}, \orgname{University of Southampton }, \orgaddress{\street{University Road}, \city{Southampton}, \postcode{SO17 1BJ}, \state{Hampshire}, \country{United Kingdom}}}



\maketitle

\section{Fabrication}\label{sec1}
The fabrication process begins with the Low Pressure Chemical Vapor Deposition (LPCVD) deposition of a 400~nm thick $\text{Si}_3\text{N}_4$ layer on a silicon substrate with a 3.2~$\mu$m thermal oxide undercladding. Waveguide geometries are patterned using high-resolution 248~nm Deep UV lithography and subsequently defined through reactive ion etching (RIE). Following core definition, a $\text{SiO}_2$ cladding is deposited and planarized using chemical mechanical polishing (CMP) to achieve the target 20~nm optical spacer thickness above the waveguide surface. The resistive microheaters are then formed using an 80~nm thick indium tin oxide (ITO) film deposited via physical vapor deposition (PVD). An annealing process follows to improve the ITO conductivity\cite{Lotkov2022, rajendran2024indium}. Electrical contact is established through the deposition of Ti/TiN/Al/Ti/Ni (top to bottom) pads. The active 20~nm thick $\text{Sb}_2\text{Se}_3$ layer is integrated via radio-frequency (RF) magnetron sputtering. To prevent degradation of the PCM during the followup fabrication steps, a 40~nm thick $\text{ZnS:SiO}_2$ (20\%:80\%) protection cladding is deposited immediately following the PCM growth. Finally, a conformal $\text{SiO}_2$ oxide cladding layer is deposited over the entire device stack, encapsulating the $\text{Sb}_2\text{Se}_3$ layer, ITO micro-heaters, and waveguide structures. This oxide overcladding suppresses oxygen adsorption at the ITO surface, thereby mitigating resistivity degradation of the heater \cite{ryzhikov2003anomalous, li2020high}.

\section{Electrical characterization of the normalized resistivity of a pair of ITO filaments}\label{sec2}
A Keithley 2401 SourceMeter was used to apply a DC voltage to a pair of ITO heaters through four electrical probes with needle tips. The probes were connected to the Keithley instrument via Bayonet Neill–Concelman (BNC) cables. A Python script was used to control the SourceMeter and record the current data.

\begin{figure}[htbp]
\centering\includegraphics[width=8cm]{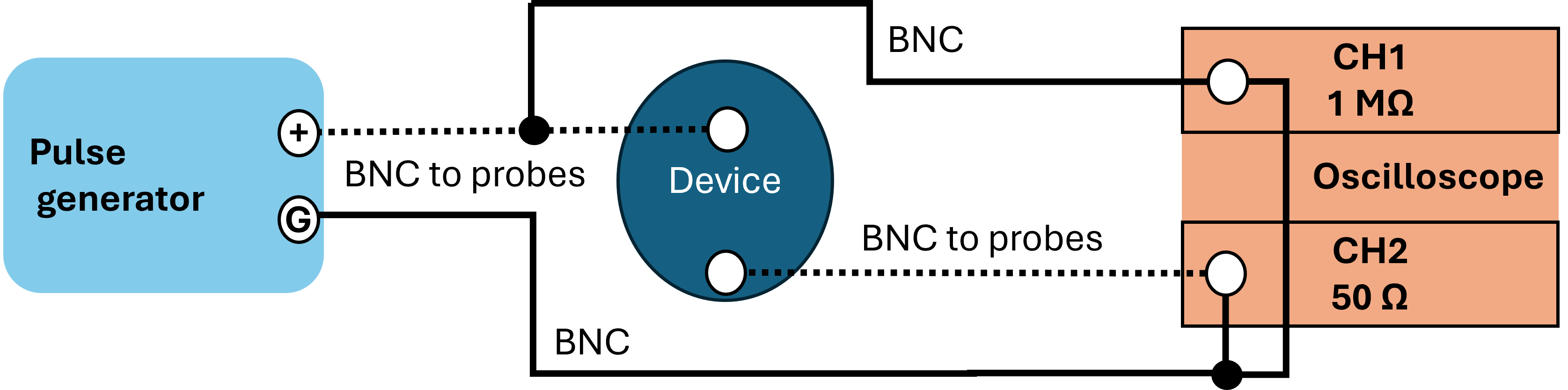}
\caption{Experiment instruments connection for heater measurement.}
\label{heater_setup}
\end{figure}

The normalized resistivity of a pair of ITO filaments in the PCM phase shifter was characterized using a pulsed electrical measurement scheme. Figure~\ref{heater_setup} shows the connection of the instruments. A pulse generator (PG, BNC577) was used to apply a single electrical pulse with peak voltage $U_1$ and pulse width $T_1$.

The output pulse was applied to a series circuit consisting of the PCM phase shifter and a 50-$\Omega$ load provided by channel~2 of an arbitrary waveform generator (AWG, Agilent 33600A). The PCM phase shifter contains two parallel ITO filaments acting as the resistive heater.

The voltage generated by the pulse generator ($U_1$) and the voltage across the 50-$\Omega$ load ($U_2$) were monitored using channel~1 and channel~2 of the AWG, respectively. Since the input impedance of channel~1 (1~M$\Omega$) is much larger than the resistance of the device and the 50-$\Omega$ load, its influence on the circuit can be neglected.

The current flowing through the circuit was determined from the voltage drop across the 50-$\Omega$ load according to

\begin{equation}
I_1 = \frac{U_2}{50\,\Omega}.
\end{equation}

The instantaneous electrical power delivered to the PCM phase shifter was calculated as

\begin{equation}
P_{\mathrm{peak}} = (U_1 - U_2) I_1 .
\end{equation}

The effective resistance of the device was obtained from

\begin{equation}
R_{\mathrm{device}} = \frac{U_1 - U_2}{I_1},
\end{equation}

which includes the resistance of the ITO filaments and the contact resistance of the metal pads ($R_{\mathrm{device}} = R_{\mathrm{ITO}} + R_{\mathrm{pad}}$). By subtracting the previously calibrated pad resistance, the normalized resistance of the pair of ITO filaments was extracted as a function of the applied peak power.

\section{COMSOL Joule-Heating Model}\label{sec3}

The temperature used for Arrhenius analysis is obtained from a finite-element Joule-heating model implemented in COMSOL Multiphysics. To validate the reliability of the extracted activation energy, an independent experimental calibration of the device temperature is performed using the thermo-optic response of a MZI. The device operates at a wavelength of 1550~nm, with an ITO heater length of 92~$\mu$m.

The COMSOL model incorporates the full device geometry (see main tex Fig.1(e)), with material parameters summarized in Table~\ref{comsolmaterial}, including electrical conductivity, thermal conductivity and heat capacity for all constituent layers (Si$_3$N$_4$, ITO, PCM and ZnS:SiO$_2$). A non-uniform tetrahedral mesh is employed, with local refinement in the heater region down to a minimum element size of 10~nm to accurately resolve steep thermal gradients.

Under DC bias below the phase-transition threshold, the measured optical phase shift is assumed to be dominated by the thermo-optic effect. The phase shift $\Delta \phi$ in the MZI is related to the effective refractive index variation $\Delta n_{\mathrm{eff}}$ as

\begin{equation}
\Delta \phi = \frac{2\pi}{\lambda}  \Delta n_{\mathrm{eff}} L
\end{equation}

where $L$ is the interaction length and $\lambda$ is the operating wavelength. The refractive index variation is in turn related to the temperature rise via the thermo-optic coefficient,

\begin{equation}
\Delta n_{\mathrm{eff}} = \frac{d n_{\mathrm{eff}}}{dT}  \Delta T
\end{equation}

where ${d n_{\mathrm{eff}}}/{dT}$ denotes the effective thermo-optic coefficient of the waveguide. Here, a value of $2.57 \times 10^{-5}~\mathrm{K}^{-1}$ is adopted, consistent with reported values for Si$_3$N$_4$ waveguides at 1550~nm~\cite{pruessner2023enhanced}. Combining the above expressions yields

\begin{equation}
\Delta T = \frac{\Delta \phi , \lambda}{2\pi L (d n_{\mathrm{eff}}/dT)}
\end{equation}

This analysis assumes that the phase response is dominated by the thermo-optic effect in Si$_3$N$_4$. The extracted temperature therefore represents an effective spatial average along the interferometer arm, while the actual temperature distribution may be non-uniform.

\begin{figure}[htbp]
\centering\includegraphics[width=8cm]{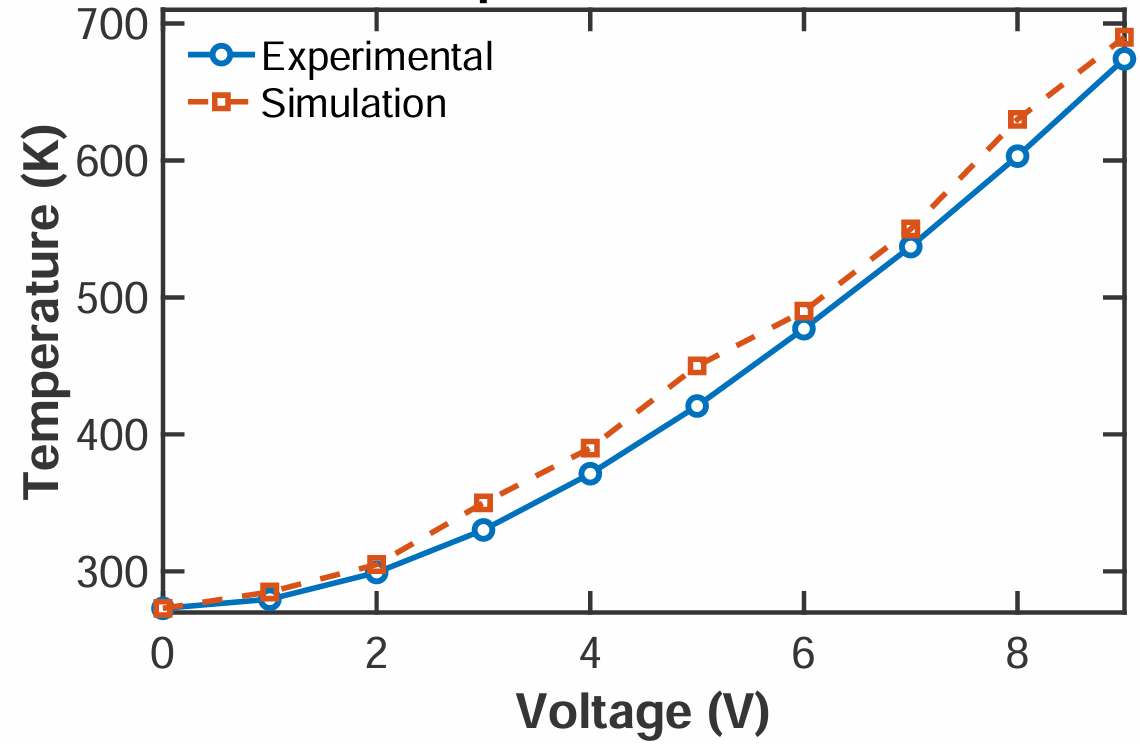}
\caption{Comparison between simulated temperature and experimentally extracted temperature from optical phase shifts.}
\label{comsolT}
\end{figure}

In Fig.~\ref{comsolT}, the experimentally extracted temperature is compared with the simulated temperature obtained under identical electrical bias conditions. The corresponding electrical power predicted by the model is also benchmarked against experimental measurements. Excellent agreement is observed over the full voltage range (0–9~V), with a mean relative error of 0.00337 and a coefficient of determination of $R^2 = 0.9970$, confirming the accuracy of the electrothermal model.

\begin{table*}
\centering
\caption{Thermal material properties used in the COMSOL model.}

\begin{threeparttable}
\resizebox{\textwidth}{!}{
\begin{tabular}{lccc}
\toprule
Material & Thermal Conductivity (W/m$\cdot$K) & Density (kg/m$^3$) & Heat Capacity $C_p$ (J/kg$\cdot$K) \\
\midrule
Si$_3$N$_4$\cite{pierson1999handbook} &  30 & 3180 & 670 \\
ITO & 4 \cite{wang2018potential} & 7140 \cite{yoo2017thermally} & 340 \cite{yoo2017thermally}\\
$\mathrm{Sb}_2\mathrm{Se}_3$ & 0.72 \cite{aryana2023optical}& 5840 \cite{aryana2023optical}& 270 \cite{pashinkin2008heat}\\
ZnS:SiO$_2$~$^{a}$& 0.6 \cite{tsu2006mechanism}& 2600 \cite{kim2000heat}& 700 \cite{kim2000heat} \\
\bottomrule
\end{tabular}
}

\begin{tablenotes}
\small
\item[$^{a}$] Estimated using a volumetric weighted average.
\end{tablenotes}

\end{threeparttable}
\label{comsolmaterial}
\end{table*}

\section{Optoelectronic cycling and transmission measurements}\label{sec4}
\begin{figure}[htbp]
\centering\includegraphics[width=8cm]{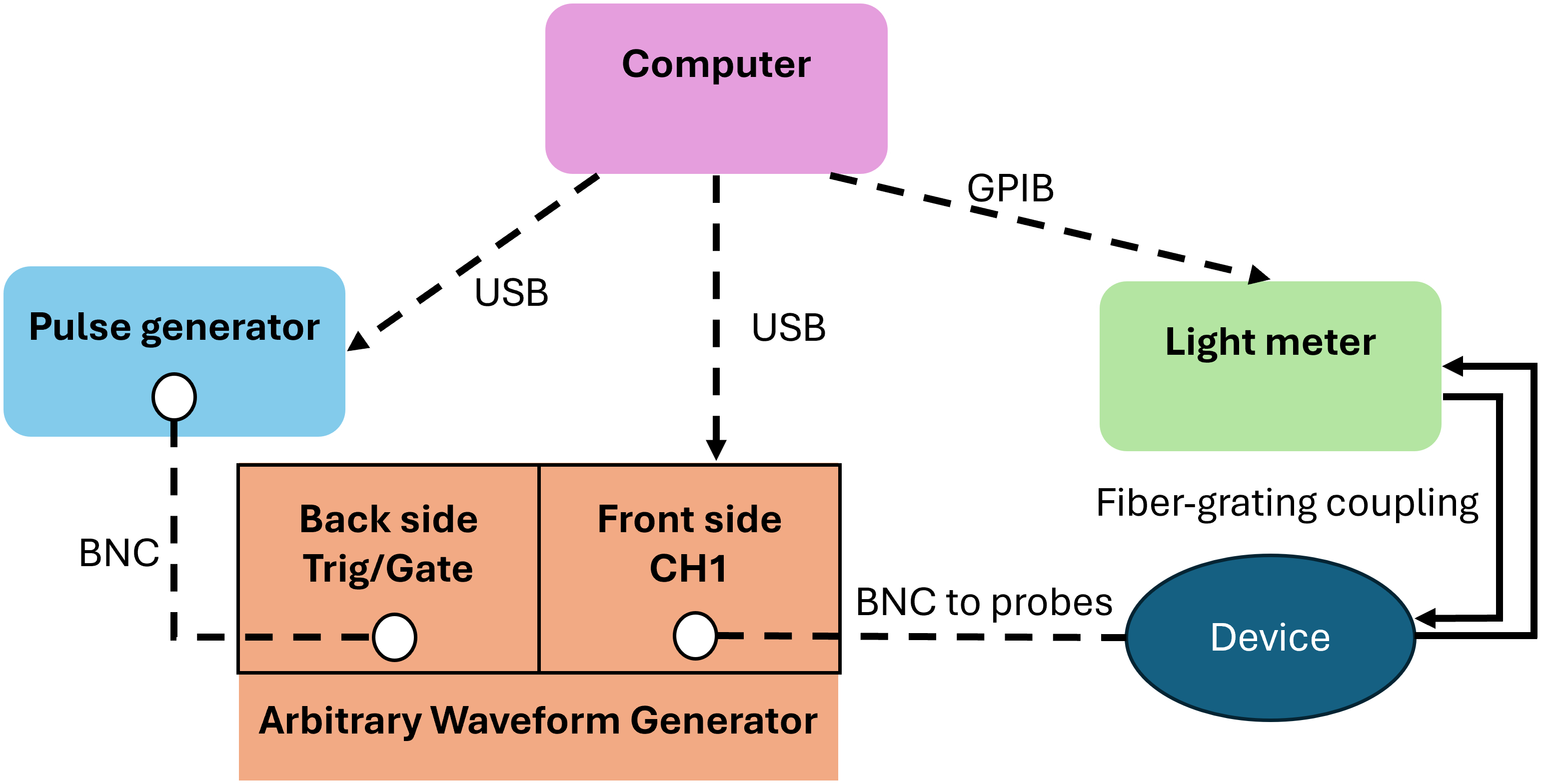}
\caption{Experiment instruments connection for cycling.}
\label{cycling_setup}
\end{figure}

The optoelectronic cycling experiment was performed on a MZI incorporating a PCM phase shifter in each arm. The PCM was initially in the crystalline state. All instruments were controlled using a custom Python program. The instrument connection is shown in Fig.~\ref{cycling_setup}.

The optical transmission from MZI input~1 to output~1 was characterized using a lightwave multimeter (Agilent 8163B). A tunable laser module (81940A) provided wavelength sweeps across the C-band, while the transmitted optical power was recorded using a calibrated power sensor (81634B). 

To accelerate the cycling experiment, electrical pulses were generated using an arbitrary waveform generator (AWG, Agilent 33600A), which was directly connected to the PCM phase shifter. It is capable of delivering voltage pulses with amplitudes up to 10~V. The rise and fall times of the pulses are approximately 100~ns. The AWG produced a periodic waveform with a period of 6~ms, consisting of a set pulse (0.006~W/$\mu$m, 20~\textmu s) and a reset pulse (0.035~W/$\mu$m, 600~\textmu s). Since the AWG does not provide an internal pulse counter, a pulse generator (PG, BNC577) was used to trigger and gate the AWG output, enabling accurate control of the total number of electrical pulses applied to the device.

After reaching a predefined number of electrical cycles, the optical characterization sequence was triggered. The AWG first applied an additional set pulse to amorphize the PCM. After a waiting time of 10~ms to ensure thermal relaxation, the laser performed a wavelength sweep and the optical transmission spectrum was recorded. Subsequently, a reset pulse was applied to recrystallize the PCM, followed by a second optical transmission measurement. The phase shift between these two switching events can be extracted from the spectral shift between the corresponding transmission minima, denoted as $\Delta \lambda$:

\begin{equation}
\Delta \lambda = \lambda_{\mathrm{reset}} - \lambda_{\mathrm{set}},
\end{equation}

The phase shift $\Delta \phi$ can be expressed as
\begin{equation}
\Delta \phi = 2\pi \frac{\Delta \lambda}{\mathrm{FSR}},
\end{equation}
where $\mathrm{FSR}$ is the free spectral range of the interferometer.

Accordingly, the phase-shifting efficiency $\eta$ is defined as
\begin{equation}
\eta = \frac{\Delta \phi}{L} = \frac{2\pi}{L} \frac{\Delta \lambda}{\mathrm{FSR}},
\end{equation}
where $L$ is the length of the PCM section.

Using this automated procedure, approximately $2\times10^5$ electrical switching cycles and one pair of set/reset optical transmission measurements could be completed within about 22~min.

\section{C band device}\label{sec5}
The static switching characteristics of the $\mathrm{Sb}_2\mathrm{Se}_3$-integrated MZI were characterized using the instrumentation described in Supplementary Section 4. Fig.~\ref{initial}(a) shows the transmission spectra of the device in the post-fabrication crystalline state, as well as after subsequent amorphization and crystallization operations. In the initial crystalline state, the MZI exhibits a well-defined interference pattern with an extinction ratio of $33~\mathrm{dB}$ at a resonance wavelength of $1545.12~\mathrm{nm}$ and a free spectral range (FSR) of $10.3~\mathrm{nm}$ in the C-band.

Upon application of a high-energy pulse ($7~\mathrm{V}$, $20~\mu\mathrm{s}$), the $\mathrm{Sb}_2\mathrm{Se}_3$ undergoes a transition to the amorphous state, inducing a substantial change in the effective refractive index and resulting in a blue shift of $1.74~\mathrm{nm}$. The extinction ratio (ER) increases slightly from 33 to $34~\mathrm{dB}$, indicating negligible imbalance between the interferometer arms. Meanwhile, the total insertion loss remains nearly constant at $1.5~\mathrm{dB}$, confirming the low-loss nature and stability of the device during switching.

To evaluate reversibility, the PCM was recrystallized using a lower-energy pulse ($5.5~\mathrm{V}$, $600~\mu\mathrm{s}$), resulting in a red shift of approximately $1~\mathrm{nm}$. This corresponds to a $\sim$57\% reduction in phase shift compared to the initial amorphization from the post-fabrication state. For this switching cycle, the maximum ER is approximately 24 dB at 1543.4 nm.

We further investigated the dependence of phase modulation on the PCM length. As shown in Fig.~\ref{initial}(b), a clear linear relationship between the induced phase shift and PCM length is observed for C-band MZIs, yielding a maximum phase-shifting efficiency of $0.022~\mathrm{rad/\mu m}$. This value exceeds the maximum phase reported in the Operational Boundaries section of the main text, as the data presented here are obtained from devices with lengths ranging from 10 to 90~$\mu\mathrm{m}$, whereas the mapping data are collected from a device of a single length. This discrepancy arises from statistical variations across multiple devices as well as device-to-device variability.

To quantify the contribution of the PCM phase shifter to the overall insertion loss of the MZI, standalone C-band $\mathrm{Sb}_2\mathrm{Se}_3$ phase shifters were characterized. The propagation losses were extracted as $0.021~\mathrm{dB}\,\mu\mathrm{m}^{-1}$ in the crystalline state and $0.018~\mathrm{dB}\,\mu\mathrm{m}^{-1}$ in the amorphous state. Additional measurements for O-band devices are provided in Supplementary Section~6.

A significant variation in phase-shifting efficiency was initially observed between the post-fabrication crystalline state and the first electrically induced recrystallized state. Identical amorphization (set) and crystallization (reset) pulse profiles were subsequently applied in all following cycles. With continued cycling, as shown in Fig.~\ref{initial}(c), the resonance shift gradually decreased and eventually stabilized at approximately $0.004~to~0.005\,\mathrm{rad/\mu m}$ after about $10^{6}$ cycles. The extinction ratio of this 2-bit switching is stable approximately at 6 dB.

This behavior arises because, following fabrication, photonic devices typically undergo an initial stabilization process before reaching reliable operation \cite{coakley2022understanding, obeng2016towards}. The atomic configuration and interfacial conditions of the PCM layer can deviate from an ideal equilibrium state due to residual stress, surface defects, and deposition non-uniformity. Consequently, during the initial switching cycles, device performance, including crystallization kinetics, amorphisation efficiency, and phase modulation depth can exhibit pronounced fluctuations. With repeated set–to-reset cycling, the PCM microstructure progressively evolves toward a quasi-equilibrium state, in which defect populations and atomic vacancies approach a dynamic balance, internal stresses are partially relaxed, and phase boundaries become increasingly well defined and reproducible \cite{kim2019phase, constantin2022new}.

\begin{figure}[htbp]
\centering\includegraphics[width=13cm]{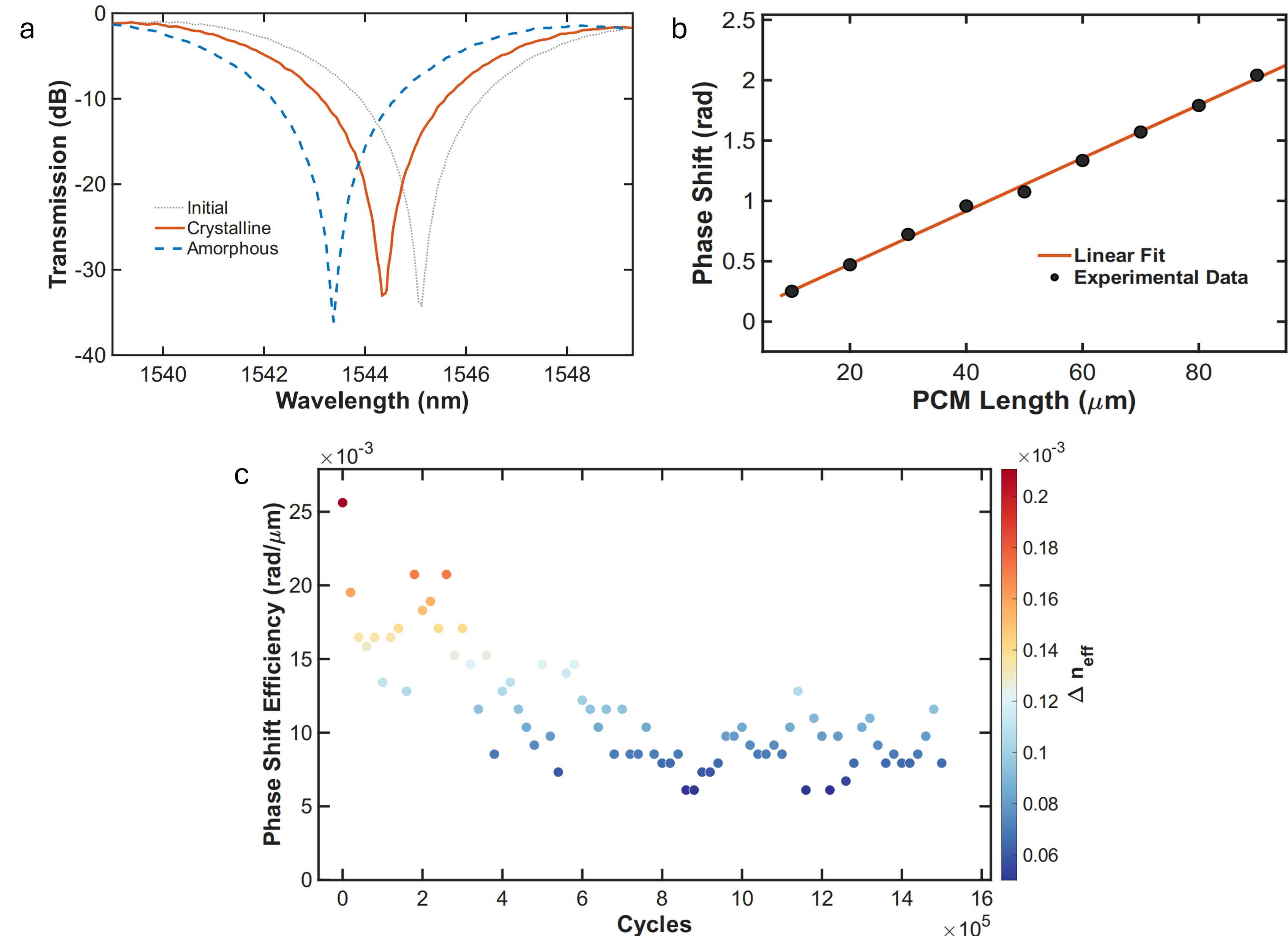}
\caption{Measurements for C band MZIs.
(a) Transmission shift at initial one cycle for a C-band MZI with $30\mu\text{m}$ PCM phase shifters.
(b) Linear fit of maximum phase change versus PCM length in C-band MZIs.
(c) Phase change in initial one million cycles for a MZI with $30\mu\text{m}$ PCM phase shifters.
}
\label{initial}
\end{figure}

\section{O band device}\label{sec6}
\begin{figure}[htbp]
\centering\includegraphics[width=13cm]{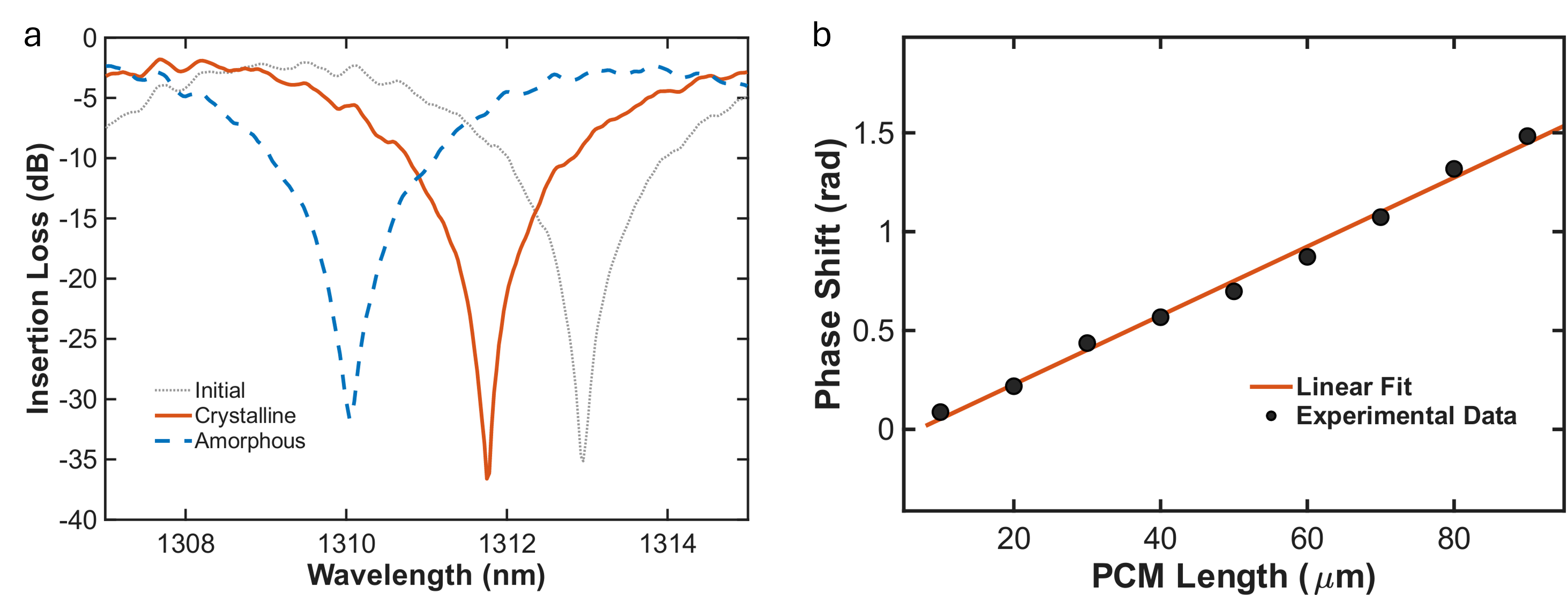}
\caption{Measurements for O band MZIs.
(a) Transmission shift at initial one cycle for an O-band MZI with $90\mu\text{m}$ PCM phase shifters.
(b) Linear fit of maximum phase change versus PCM length in O-band MZIs.
}
\label{Oband}
\end{figure}

Using the same characterization methodology as for the C-band devices, we investigated the initial switching behaviour of O-band MZIs with varying PCM lengths. Figure~\ref{Oband}(a) shows the evolution of the transmission spectrum from the post-fabrication state to the first set (amorphous) state and subsequently to the first reset (crystalline) state. A similar behaviour to that observed in the C-band devices is obtained. The total insertion loss of the device is approximately $2~\mathrm{dB}$, while the extinction ratio is around $30~\mathrm{dB}$ for all three states.

Figure~\ref{Oband}(b) presents a linear fit of the maximum phase shift as a function of PCM length, yielding a phase-shifting efficiency of $0.017~\mathrm{rad}\,\mu\mathrm{m}^{-1}$. The key performance metrics of both O-band and C-band devices are summarized in Table~\ref{tab:properties}.

\begin{table}[htbp]
\caption{Characterization of MZI Devices at Initial Cycles}
  \label{tab:properties}
  \centering
\begin{tabular}{ccccc}
\hline
Wavelength & 2-bit ER ($dB$) & Max-ER ($dB$) & Shift efficiency ($rad/\mu\text{m}$) & Total IL ($dB$) \\
\hline
O band & 25 & 30 & 0.017 & 2\\
C band & 29 & 33 & 0.022 & 1.5\\
\hline
\end{tabular}
\end{table}

\section{Amplify of set and reset pulses during 140 million switching}\label{sec7}
\begin{figure}[htbp]
\centering\includegraphics[width=13cm]{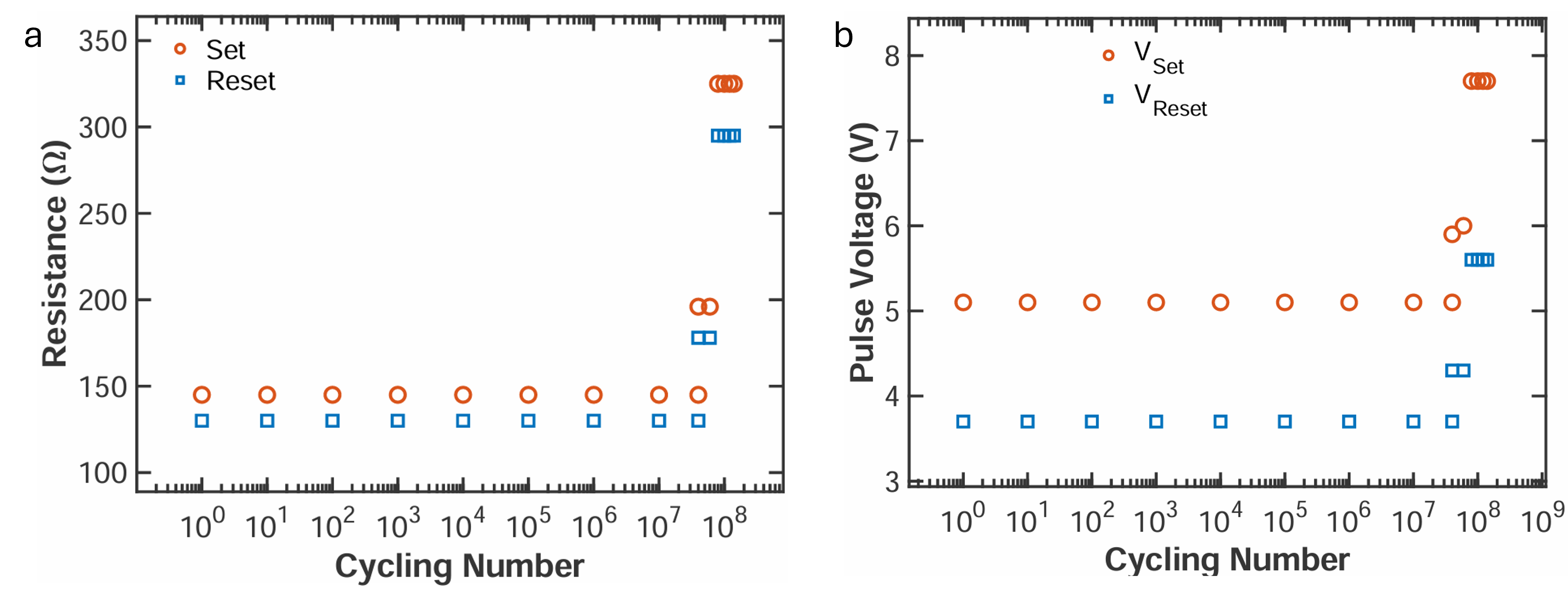}
\caption{ITO filament performance during 140 million switching cycles.
(a) Resistivity of the ITO filament under set and reset pulses over 140 million cycles.
(b) Applied voltages of the set and reset pulses on the ITO filament over 140 million cycles.
}
\label{140ITO}
\end{figure}

This section presents the performance of ITO heaters over 140 million switching cycles. The resistivity increases by approximately 40\% after 40 million cycles and further rises after 80 million cycles, as shown in Fig.~\ref{140ITO}(a). Figure~\ref{140ITO}(b) shows the increase in the amplitude of the set and reset pulses required to maintain the peak power for consistent switching conditions.

\bibliography{sn-bibliography}